\documentclass{article} 
\usepackage[utf8]{inputenc}
\usepackage{graphicx} 
\usepackage{geometry}
\usepackage{amsmath}
\usepackage{authblk}
\usepackage{natbib}
\usepackage{color}
\usepackage{easyReview}
\usepackage{algorithm}
\usepackage{algpseudocode}
\newcommand{\base}{2}
\newcommand{\Podo}{\textit{P.~anserina}}

\renewcommand{\add}{}

\title{Full identification of a growing and branching network's spatio-temporal structures.}

\author[1]{Thibault Chassereau}
\author[1]{Florence Chapeland-Leclerc}
\author[1,*]{\'Eric Herbert}
\affil[1]{Université Paris Cité, CNRS, UMR 8236-LIED, F-75013 Paris, France}
\affil[*]{Corresponding author : eric.herbert@u-paris.fr}
\date{ Compiled on \today }

\begin{document}
\maketitle
\begin{abstract}
Experimentally monitoring the kinematics of branching network growth is a tricky task, given the complexity of the structures generated in three dimensions. One option is to drive the network in such a way as to obtain two-dimensional growth, enabling a collection of independent images to be obtained. The density of the network generates ambiguous structures, such as overlaps and meetings, which hinder the reconstruction of the chronology of connections.
In this paper, we propose a general method for global network reconstruction. Each network connection is defined by a unique label, enabling it to be tracked in time and space. In this work, we distinguish between lateral and apical branches on the one hand, and extremities on the other. Finally, we reconstruct the network after identifying and eliminating overlaps. 
This method is then applied to the model filamentous fungus \textit{Podospora anserina} to reconstruct its growing thallus. We derive criteria for differentiating between apical and lateral branches. We find that the outer ring is favorably composed of apical branches, while densification within the network comes from lateral branches. From this, we derive the specific dynamics of each of the two types.
Finally, in the absence of any latency phase during growth initiation, we can reconstruct a time based on the equality of apical and lateral branching collections. This makes it possible to directly compare the growth dynamics of different thalli.
\end{abstract}

\section*{Significance}

A complete description, i.e. both local and global, of a growing and branching network enables us to discuss the microscopic rules underlying emerging macroscopic phenomena. This task is made complex by the different scales involved, the high network density and network masking.
Here, we propose a dynamics-based reconstruction method derived from a collection of bright-field images. We obtain a unique label for each structure, which can then be tracked in time and space, enabling their respective quantification during growth, and a first systematic classification of branches. These results pave the way for a better description of local rules and for realistic simulations.  

\section*{Introduction}

Networks are commonplace objects, and their study has been facilitated by the establishment of online databases \cite{nr, snapnets}. In this work, we are interested in growing spatial networks, which show a number of vertices and edges that increase over time, such as the layout of cities \cite{Lagesse2015Mar}, the propagation of cracks in drying clay \cite{Bohn2005Apr}, or the growth of a gorgonian \cite{Valcke2020Nov}. In nature, filamentous networks are ubiquitous (shoot and root systems of land plants \cite{Fernandez2022Dec}, and in animal organs such as the lung, kidney, mammary gland, vasculature, etc) and constitute a key step in the evolution of morphological complexity \cite{coudert_design_2019}. What these networks have in common is a polarized growth process (at the end of the edges), a branching mechanism (new tip) and a merging mechanism (tip deletion). Experimentally, they generally are studied from time-lapse recordings and present the difficulty of reconstructing the dynamics from a collection of independent images. 
A representative example of such a growing branching network is the thallus of a mycelial fungus.
Characterizing its growth is a tricky task. Hyphae form a complex, static yet growing network \cite{falconer_biomass_2005},  without being able to define a homogeneous region of occupation, \textit{ie} there is no clearly defined surface.
In the literature, mycelia are usually described on a macroscopic scale using quantities such as mass or apparent thallus diameter. However, the filamentous nature of the mycelium makes these quantities unsuitable. It is therefore no longer possible to distinguish the different phenomena at work during growth, which would enable us to characterize different species, stresses or mutants \cite{ledoux_characterization_2024}. For example, connection rates, or the difference in density between the centre and the periphery. 
The initiation phase was also rendered inaccessible as the thallus had barely formed and was too small for macroscopic measurement \cite{ulzurrun_fungal_2019, de_ligne_analysis_2019}. \\
At the microscopic level, studies were limited to localised manual surveys on analogue and then digital images until the 2010s, using white light \cite{boswell_functional_2002, bebber_biological_2007, trinci_hyphal_1973, trinci_study_1973,  fricker_imaging_2008,  reynaga-pena_analysis_1997} or, more rarely, radionuclides \cite{gooday_autoradiographic_1971}.
Automation of the image acquisition and processing has enabled a new experimental period to be opened up. Specific observables (such as length, orientation etc) can be quantified over time, by monitoring single hypha but the global dynamics become then inaccessible \cite{barry_morphological_2009, sanchez-orellana_automated_2018, schmieder_bidirectional_2019}.
The entire mycelium has also been studied in two dimensions \cite{vidal-diez_de_ulzurrun_automated_2015,  fricker_mycelium_2017, dikec_hyphal_2020} and three dimensions using confocal laser scanning microscopy \cite{du_morphological_2016} (see also \cite{chiang_three-dimensional_2011} that consider Brain-wide Wiring Networks in Drosophila) or  X‐ray microcomputed tomography \cite{schmideder_three-dimensional_2019}. 
These observations were used to discuss changes in overall length, exhaustive collections of free apexes, fraction of biomass and branch angles using specific software, generally based on graphical reconstructions of the network. These approaches consider collections of independent images, so it is not possible to automatically track an apex through time. 
Overlaps, and in particular the distinction with network connections, are then particularly tricky to deal with, even when relying on a 3-D reconstruction \cite{chiang_three-dimensional_2011}. 
\add{
    This is a recurring problem for the reconstruction of branching structures. For example, to discuss the root system architecture of \textit{Arabidopsis thaliana} seedlings a recent work  by \cite{Fernandez2022Dec} proposed a method to reconstruct small networks of developmental root. This work aimed to automatize the reconstruction process. To the best of our knowledge this is the only proposition of a global pipeline, from data acquisition to graph reconstruction, integrating branching and overlaps. This was made possible by relying on root growth specificities.
    Several points are crucial in graph reconstruction and need to be addressed differently in the study of thallus growth, namely connectivity, branching and orientation. 
    First, there is no process similar to fusion/anastomosis, which removes the ambiguity concerning the reconstruction of connectivity. 
    Second, the branching process remains low compared to what is observed in thallus growth. From the primary root, the most frequent studies discuss first-order lateral branching. They focus on the first 5 to 7 days of growth hardly showing higher order lateral roots as discussed in \cite{deja-muylle_genetic_2022}. 
    Third, growth is generally oriented towards exploring the surrounding space, with roots system that can interact with their neighbors and create structures such as swarms \cite{ciszak_swarming_2012}. The location of the root in relation to the primary root is a strong indicator of its age. Conversely, in the case of a thallus, the increase in density generated by the multiple branches makes the orientation and age of a specific hypha unpredictable.
}

Basing our analysis on the kinematics of the network's growth, we propose in this paper a complete reconstruction method that associates a unique identifier with each node in the network across time and space. It then becomes possible to automatically track the exhaustive collection of network apexes over time.
We will focus our analysis on \textit{Podospora anserina}~\cite{silar_podospora_2020}, a coprophilous filamentous ascomycete, a large group of saprotrophic fungi, that mostly grows on herbivorous animal dungs and plays an essential role within this complex biotope in decomposing and recycling nutrients from animal feces. Then, \Podo{} can be considerred as a short-range forager, displaying usually higly dense hyphae that are successful in encountering small organic food resources \cite{sanati_nezhad_cellular_2013}. \Podo{} has long been used as an efficient laboratory model to study various biological phenomena, especially because it rapidly grows on standard culture medium, it accomplishes its complete life cycle in only one week, leading to the production of ascospores, and it is easily usable in molecular genetics, cellular biology and cytology. 
\Podo{} is a model for studying the growth of biological networks, as the its thallus shows great complexity, but its growth follows a simple set of rules.  We recall some details concerning the thallus growth of \Podo{}: (i) growth is highly polarized, \textit{ie} it occurs only through the apex of the hyphae, (ii) once formed, the hyphae are immobile, (iii) a new hyphae is created during the branching process, which generates a new apex. It can be located near an existing apex or elsewhere on the network (iv) when two hyphae meet they may fuse (anastomosis).

In this work, we build on a graph structure to describe this network, since these rules can be easily translated.   First, the initial spore is the only permitted origin of the network. The network is static ; \textit{ie} hyphae grow in length but do not move or disappear. The network is considered to grow only in 2D ; the only vertical movement permitted is overlap. Hyphae are always in contact with the substrate. The temporal dynamics of growth allow causality (the network must be contiguous).

We extract the information needed for a complete description, in space and time, of the network growth. We identify and label each biologically relevant vertex (also called node) to enable individual temporal tracking of each vertex, disambiguating the different types of crossover based on the spatio-temporal dynamics of the growing network. Building on the vertex collections we've built, we then quantify each type of branching over time. Our method allows to 
(i)--label each network vertex with a unique identifier, enabling it to be tracked over time. In particular, we plan to indicate the location and time at which each vertex appeared. 
(ii)--At each time step, the growth of each hypha, \textit{ie} the displacement of each apex, is measured. Each apex is identified in time and space and connected to the correct branch. Lateral branches are distinguishable from apical branches. The direction of growth is regained. 
(iii)--Distinguish between branchings (connections), anastomoses (fusions) and overlaps for each degree-3 vertex. Overlap are eliminated from the network connections list and the corresponding network reconstructed accordingly. Finally, complete hyphae, from the branching to the apex, are reconstructed. 

\section*{Materials and methods}

In this section we will briefly discuss the \textit{experimental setup}, largely described elsewhere. Then we describe the reconstruction process of the network spatio-temporal structure. In the subsection \emph{Temporal coordinate} we will see how a temporal coordinate is set to each node in the network. The subsection \emph{Branch identification and twig's orientation} will focus on how we define and identify each branch in the network. The last subsection \emph{Degree-3 vertices classification and overlaps} will deal with the classification of each degree-3 vertices according to its biological nature and the detection of overlaps. \add{This pipeline is illustrated in figure \ref{fig:recap}. A pseudo-code version of the various stages of the reconstruction is available in the appendix \ref{annexe:PseudoCodes}.}

\begin{figure}
	\includegraphics[width=\linewidth]{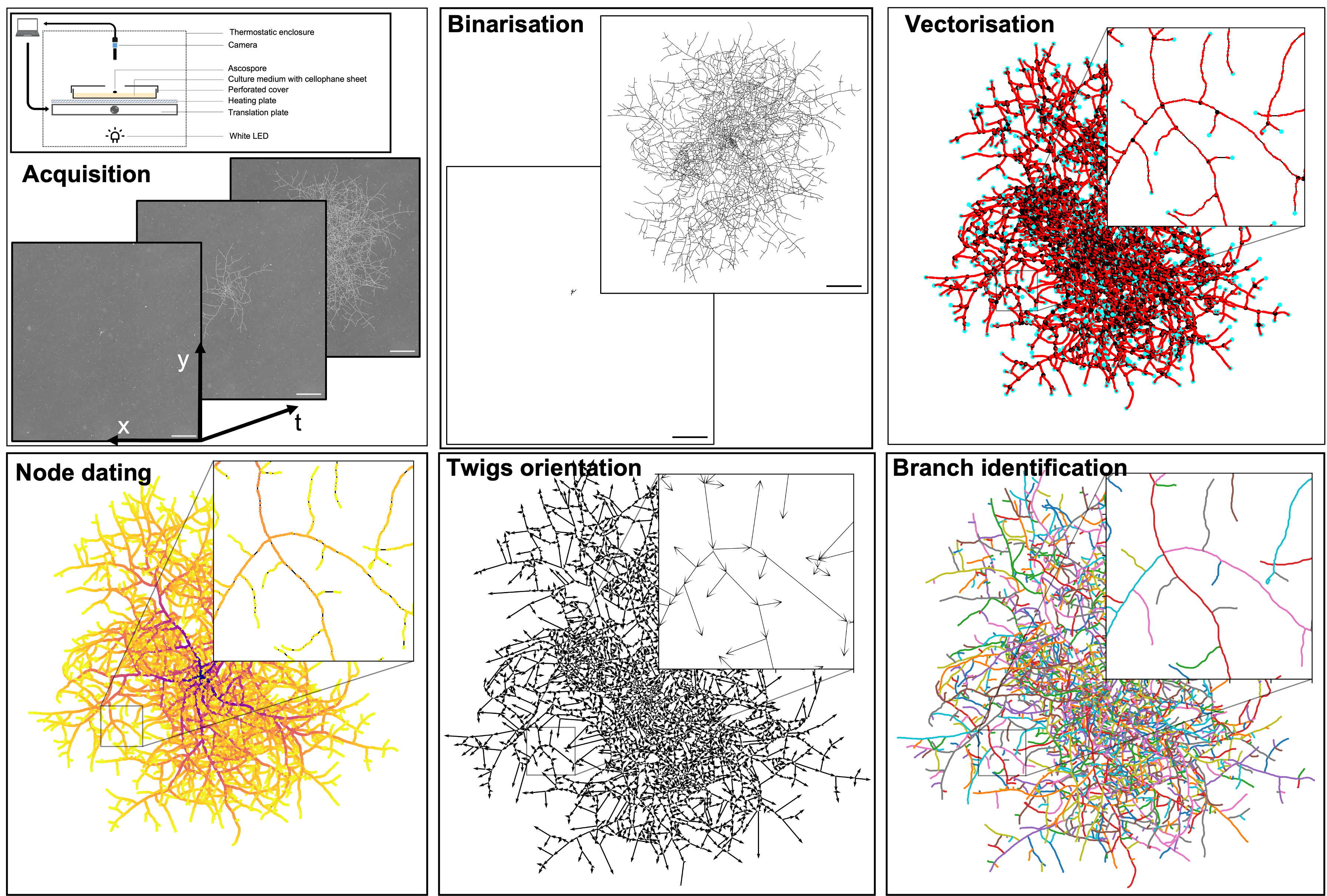}
	\caption{\textbf{Summary of the network reconstruction pipeline.} \textbf{Acquisition:} greyscale images are taken during the growth of the mycelium starting from a germinating ascospore. \textbf{Binarisation:} manual binarisation of the final and first image of the collection. \textbf{Vectorisation:} from the final binarised image, a unique spatial graph is generated allowing only degree 1 nodes (cyan), degree 2 nodes (red) and degree 3 nodes (black). \textbf{Node dating:} using grey scale images from the acquisition step and the first image binarised, each node of the spatial graph get a time coordinate. \textbf{Twigs orientation:} each twig (collection of nodes between two degree 3 nodes) is aligned according to its growth using time information and alignment with neighbouring twigs. \textbf{Branch identification:} branches are identified by following the flow indicated by twigs orientation, temporal synchronisation of branchings and spatial alignment.}
	\label{fig:recap}
\end{figure}
\subsection*{Experimental setup}

In, \cite{dikec_hyphal_2020, ledoux_prediction_2022, ledoux_prediction_2023} we discussed the global growth of hyphal network expansion and structure of \Podo{} under controlled conditions based on the temporal series of centimetric image size, with a typical micrometric resolution, of the network dynamics, starting from germinating ascospores.
An experimental device allowing to solve the dynamics of the local and global growth of the complete hyphal network of {\it P. anserina}  directly on a Petri dish from an ascospore and over a period of approximately 20 hours in a controlled environment has been previously developed and described in~\cite{dikec_hyphal_2020}.  
We did use of this setup to carry out three, named $(A)$, $(B)$ and $(C)$ thereafter, complete and independent series of a collection of greyscale images of the thallus growth, from a single initial micrometric spore to a complex network of thousands of interconnected hyphae,   under standard growth conditions with M2 culture medium, see \cite{silar_podospora_2020}  for more details, at an optimal temperature of $27^{\circ}$C. \add{Images corresponding to $(A)$, $(B)$ and $(C)$ experiments can be found online at \cite{herbert_movies_2023}.}
A sheet of cellophane is placed on top of the culture medium. The porosity of this sheet allows the nutrients to pass through and nourish the growth of the thallus and its rigidity prevents the apexes from penetrating into the nutrient medium and keeps the network at its surface. The growth of the network is then driven mostly in two dimensions during the observation period. It leads to a series of grey-scale images taken every $T = 18 $\,min with a spatial resolution of $1.6 \mu $m/pixel.
No confinement is imposed at the top of the network, allowing hyphae to cross each other by overlapping. 
As the camera is fixed above the thallus, the images obtained are the projection of the entire network on this axis. Thus, the connections observed on the network images may be real, \textit{ie} composed of branches or anastomoses (fusion of hyphae) or artifacts, \textit{ie} overlaps. The distinction between branches, anastomoses and overlaps is then made impossible, preventing the simple numeration of significant biological objects, the branches, apart from an approximate accounting, as proposed in~\cite{dikec_hyphal_2020}.  A crucial part of the work presented in this article is to propose a method of disambiguation of the different connection types.
\add{Images are composed of numerous tiles merged together.}
The standard binarisation and vectorization process described in~\cite{dikec_hyphal_2020} allowed for extracting the associated digital network independently for each image. \add{The successive growth images were vectorized independently for each time step. For this purpose, we used an adaptation of the library proposed in~\cite{lasser_net_2017}, based on a triangulation method to determine the centre of mass of each hyphal segment.} The objects making up the network are not identified identically in each image, which means they cannot be tracked individually over time. These digital networks correspond to graphs made up of degree\,1 vertices (corresponding to the apexes), degree\,2 vertices (corresponding to the hyphal skeleton) and degree\,3 vertices (corresponding to branches, mergers, encounters and overlaps). Vertices with more than 3 connections are rejected as having no biological significance. \add{Therefore, overlaps will correspond to two vertices of degree\,3, as can be seen in the summary picture \ref{fig:recap}.}

\subsection*{Temporal coordinate}
\begin{figure}
    \centering
    \includegraphics[width=\linewidth]{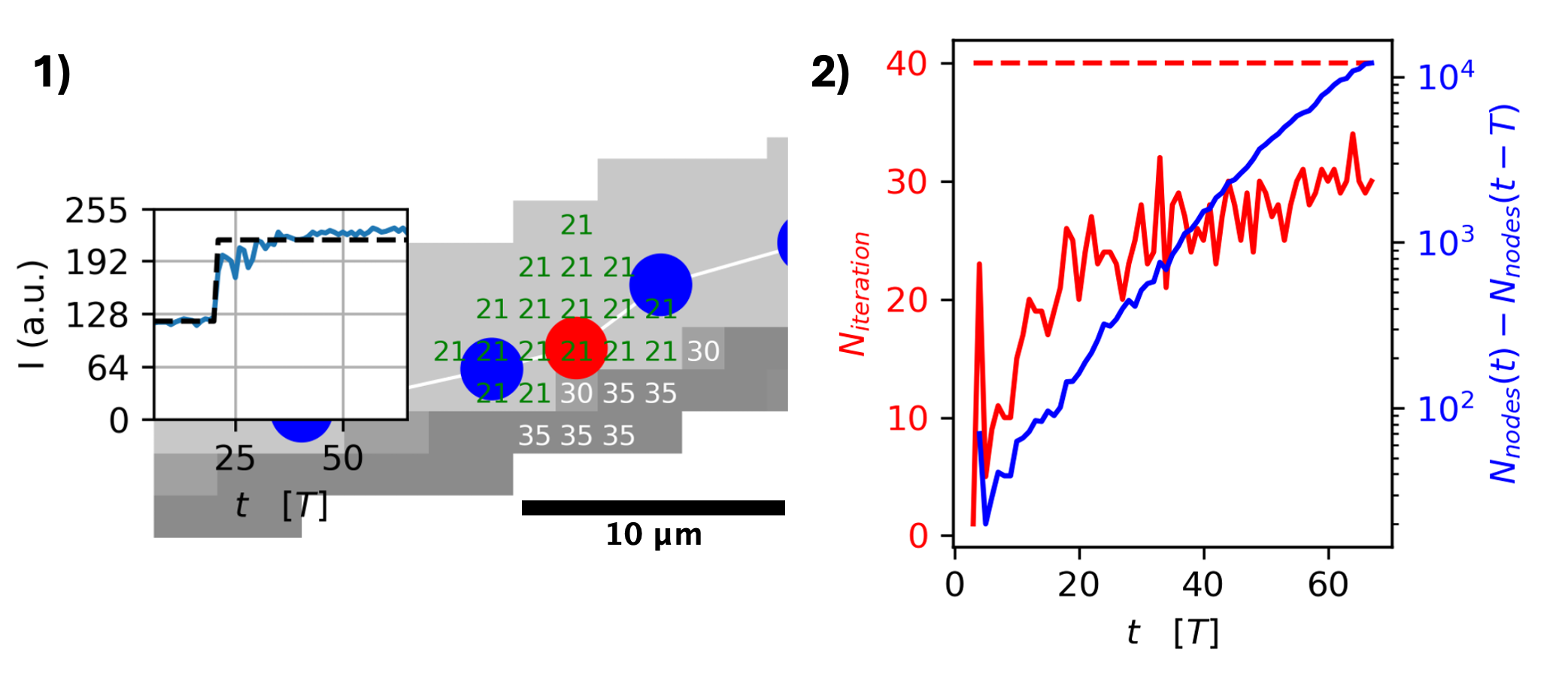}
    
    \caption{\add{\textbf{Estimation of the vertex time coordinate.}} \add{\textbf{1)} Example of time coordinate estimation of a node of interest here shown in red. The gray level of a pixel over time (blue line in inset) is fitted (dashed line in inset) using a step function (see Equation \ref{eq:StepFunction}). By reproducing this operation in the vicinity of the red pixel (defining the area of interest), one obtain the corresponding collection of time coordinates $t_0$ (see numerical values in the main figure). The time coordinate of the node of interest is the minimum (shown in green) of this collection.  Scale bar: 10 $\mu m$.
    \textbf{2)} The resulting graph must be connected at all times. Therefore the graph at time $t$ is obtained from the graph corresponding to the previous frame by adding only the neighboring vertices that satisfy $t_0 \leq t$. These new vertices are added iteratively until the current apex position is reach for each hyphae. Here we show the number of such iterations (in red, left axis) and the total number of new nodes added (in blue, right axis) at each time step. The horizontal red dashed line corresponds to the $N_{iteration}$ threshold above which a potential issue is detected.}}
    \label{fig:temporalEstimate}
\end{figure}

\textbf{Direct vertex dating}.
As the hyphae of \Podo{} can only grow and not move, every point in the network present at a time $t_1$ is present for all $t_2\geq t_1$. 
Consequently we derived only one graph of the network from the last image of our experiment. Each hypha is then allowed to grow only in this network. Vertex location is determined by the final network graph. \add{In accordance with \cite{Fernandez2022Dec}, growth is understood as the time at which each node appears. }
Therefore the goal here is to find, for each node of the network, at which time $t_0$ the node is present for the first time. As in our setup,  hyphae appear white on a black background, $t_0$ is calculated from the evolution over the time $t$ of the luminosity intensity $I$ of the pixels around the node. Theoretically, we expect to see a step function with the following expression:
\begin{equation}
    S_{t_0}(t) = \left \{
    \begin{array}{lrl}
         I_{background} & \text{if} & t<t_0\\
         I_{hyphae} & \text{if} & t\geq t_0
    \end{array} \right .
    \label{eq:StepFunction}
\end{equation}
For each pixel that is activated in the final binarized image, we find the best fit for the parameter $t_0$ through least square method, see figure\,\ref{fig:temporalEstimate} \textbf{1)}. $I_{background}$ and $I_{hyphae}$ are chosen to be the mean gray level value respectively before and after $t_0$. For each pixel around the node, we find the value for $t_0$ that minimize the deviation from the step function. The first estimation of $t_0$ for the node is then the minimal value for this parameter among those pixels. This selection is set to take all pixels within half a hyphal diameter of the node (a hypha is estimated to be 7 pixels wide). This method requires good contrast of luminosity between the hyphae and the background. 

The vast majority of nodes can be dated in this way. Locally, however, binarization may have been biased by the presence of impurities or a loss of focus. This leads to two problems, which we address in turn below. The correction of a poor estimate of $t_0$ and the correction of a hyphae discontinuity. \\
\textbf{Correction of wrong estimation of $t_0$}.
As all nodes are connected, an erroneous estimate of $t_0$ can simply be identified by comparing it with the $t_0$ of its nearest neighbors. Each node must be connected to an  older or of the same age node. 
Once a spurious estimate is detected for a specific node, the corresponding $t_0$ is replaced by an arithmetic mean of the nodes surrounding it. 
\add{As we look at the growth from a single ascospore, the graph should be connected at any time $t$. Therefore a final correction to $t_0$ is made. We start by using the first image binarized as a mask to identify the starting nodes of the network. We then build timestep after timestep, the corresponding subgraphs. For a given frame $t$, we start by initialising the corresponding subgraph as the one from the previous timestep. Among all the vertices that are not in the current subgraph, we add all those which are neighbors to at least one node of the current subgraph and whose estimate of $t_0$ is lesser or equal to $t$. As more than one node can be added at each apex in a single timestep, this step is repeted until no more nodes are added. We keep track of the number of this iterations needed to generate the subgraph from the previous one and it can be seen in figure \ref{fig:temporalEstimate} \textbf{2)} along with the total number of nodes added at each time.} \\
\textbf{Correction of broken edges}
In the case of impurity or loss of focus, the density contrast may be insufficient for binarization to detect the hypha. These defects appear as discontinuities. 
In our procedure, nodes are added with a continuity condition.  Breaking an edge stops the procedure, and prevents further identification of the growth.
The network features a large number of branches and crossings. So it's extremely likely that the hypha is connected via another part of the network. It will eventually be added to the network, but in a particular way, \textit{ie} the entire length will appear in a single time step. These situations are detected and manually corrected (generally by reconstructing the continuity of the incident hypha).
\noindent
The network is reconstructed from contiguous vertices. Branches thus appear spontaneously as dated vertices in a previously existing segment. 
The dating of branches is reconstructed as the instant when the length of the considered hypha is zero.
We now have the network of dated vertices. We now turn our attention to branch reconstruction.

\subsection*{Branch reconstruction and orientation}

\begin{figure}
    \centering{}
    \includegraphics[width=\linewidth]{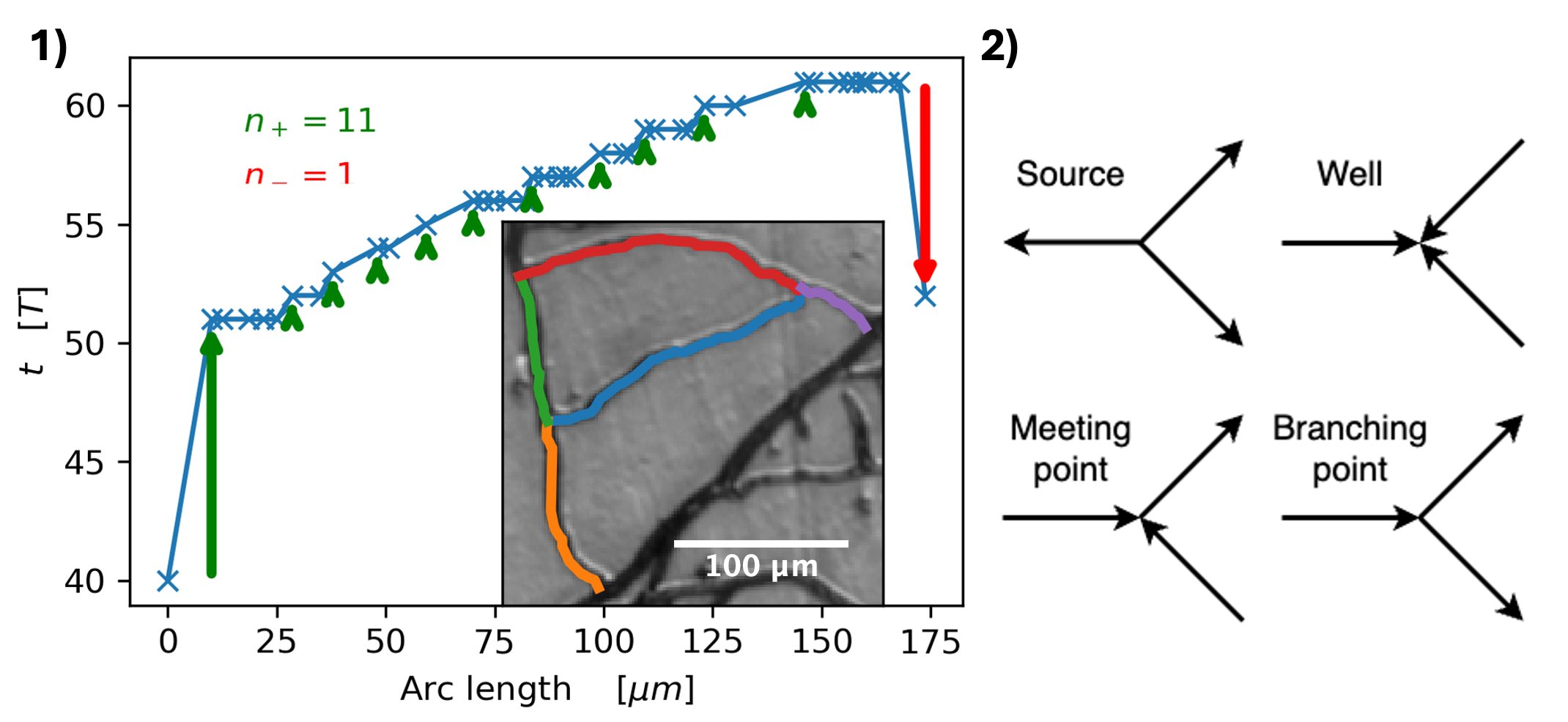}
    \caption{\add{\textbf{Determination of twig orientation.}} \textbf{1)} Example of apparition time vs the arc length for nodes along a twig. Increment and decrement are indicated by respectively green or red arrows. Inset shows the corresponding picture of the network with in blue the twig of interest \add{and its four neighbors in different colors for better identification}. \add{Scale bar: 100 $\mu m$.} \add{\textbf{2)} Nomenclature of the different possible configurations for the orientation of twigs at a degree-3 node.}}
     \label{fig:TwigsOrientation}
\end{figure}

The vertices form interconnected filaments, or hyphae, which are the structure of the network. We are now going to reconstruct each of these hyphae using the time-space coordinates of the collection of vertices defined in the previous section. 
The definition of a hypha as a long filament which collectively form the mycelium is a little vague. To avoid any confusion and in order to generalise the problem to non-fungal networks, we introduce the notion of branch, defined as the set of vertices in the network corresponding to the trajectory of an apex. In the remainder of this document, we will use the terms \textit{hyphae} and \textit{branches} interchangeably.
\noindent
At first glance, a branch is easily identified as a succession of degree-2 nodes between a degree-3 node corresponding to the initial branching of the branch and an apex represented by a degree-1 node. A branch may also terminate with a degree-3 node when it merges with another hypha. However, this branch will itself be the origin of other branches and may be the site of anastomosis of other branches.  The collection of vertices that make up the branch is therefore more varied and complex. We can think of it as a series of degree-2 vertices, each framed by vertices of degree-3 and/or 1.
 We define these series of degree-2 nodes between two degree-3 or degree-1 nodes as twigs (called terminal twigs if they end in a degree-1 node). A branch is then a succession of twigs.
\noindent
Let us now turn to the collection of twigs to be chosen to reconstruct a branch. Assuming that all the twigs are oriented in the direction of growth of the network, identifying the branches means associating locally, at each degree-3 vertex, which twigs correspond to the same branch. Each degree-3 vertex must follow one and only one of the following four rules: 
\begin{itemize}
    \item Source: three outgoing twigs. From a biological point of view, as we start from a unique spore, this source can only be unique too. Each outgoing twig is considered to be the starting point of a new branch.
    \item Wells: three incoming twigs. This case corresponds to 3 apexes in the same place at the same time. Although rare, this configuration is permitted during reconstruction. The three corresponding twigs are considered to be the terminations of their respective branches.
    \item Branching: two outgoing twigs and one incoming twig. 
    \item Meeting: one outgoing twig and two incoming twigs.
\end{itemize}
Those different cases are shown in the figure \ref{fig:TwigsOrientation}-2).

Based on the previous rules, we deduced the direction of growth of each twig incrementally. 
The boundary conditions are the collection of the direction of the terminal twigs, that can be determined without any ambiguity.
We then propagate the rule for orienting the terminal twigs recursively. The initial spore is the only one source allowed. If our network contains no encounters, all directions of all branches are found in this way.\\
However, since it is not the case, some twigs orientation remain undecidable with this single step especially in the denser center of the network.
To achieve this, we have built a second step based on a confidence level $c_k$ for each twig $k$ that depends only on the direction of the apex during the twig formation. For a given orientation we note $n_+$ the number of temporal increments (positive change in the time coordinate when traveling along the twig) and $n_-$ the number of temporal decrements (negative change in the time coordinate when traveling along the twig). The confidence level is then defined by the following equation:
\begin{equation}
    \forall k \in \{\text{twigs}\}, ~
    c_k = \frac{n_+-n_-}{n_++n_-} \,
    \tanh{\left(\frac{n_++n_-}{\mathcal{N}}\right)}
\end{equation}
Where $\mathcal{N}>0$ is a constant used to set the number of increments by which the user can really be confident about the direction of the twig. We made use of $\mathcal{N} = 3$. With this expression, $c_k$ varies between $+1$ if the initial direction  (define as the one verifying $n_+\geq n_-$) of the twig is in agreement with the growth directions of each time step and $-1$ in the opposite way. Changing the direction of growth of the twig simply corresponds to switch $n_+$ and $n_-$ and change the sign of $c_k$.
This confidence alone is sufficient to find the correct direction for twigs long enough to contain several increments, but for shorter twigs it is not adequate. \add{In fact, if a twig is small enough, it is possible that all its nodes appeared at the same time making it difficult to determine the direction of growth.} To this end, we introduce a score $s_k$ combining this confidence level and the alignment with the neighbouring twigs $\mathcal{V}(k)$ following the equation:
\begin{equation}
    \forall k \in \{\text{twigs}\}, ~  
    s_k = c_k + 
    \alpha \, (1-|c_k|) \,
    \sum_{\ell \in \mathcal{V}(k)}\textbf{u}_k\cdot \textbf{u}_\ell \, c_\ell
\end{equation}
With $\textbf{u}_{k,\ell}$ the unit direction vector of the twig and $\alpha$ a coefficient used to gauge the relative importance of alignment in relation to confidence. Typically we set $\alpha = 1/4$ as there is $4$ neighbouring twigs. This allows to give the same weight for space and time.
We then assume the best orientation of the remaining twigs is given by the maximisation of the total score of the network, defined as the sum of all local score. We achieve this by using a Monte-Carlo algorithm where we check in a random order all twigs and switch there orientation with a certain probability defined by the score's variation induced. The Monte-Carlo ``temperature" $\beta$ is set to $1$ and then doubled every hundred iterations. The process is stopped when the acceptance rate is below $1\%$ or after reaching $\beta = 1024$.

\subsection*{Degree-3 vertex classification and overlaps}

\begin{figure}
    \centering{}
    \includegraphics[width=\linewidth]{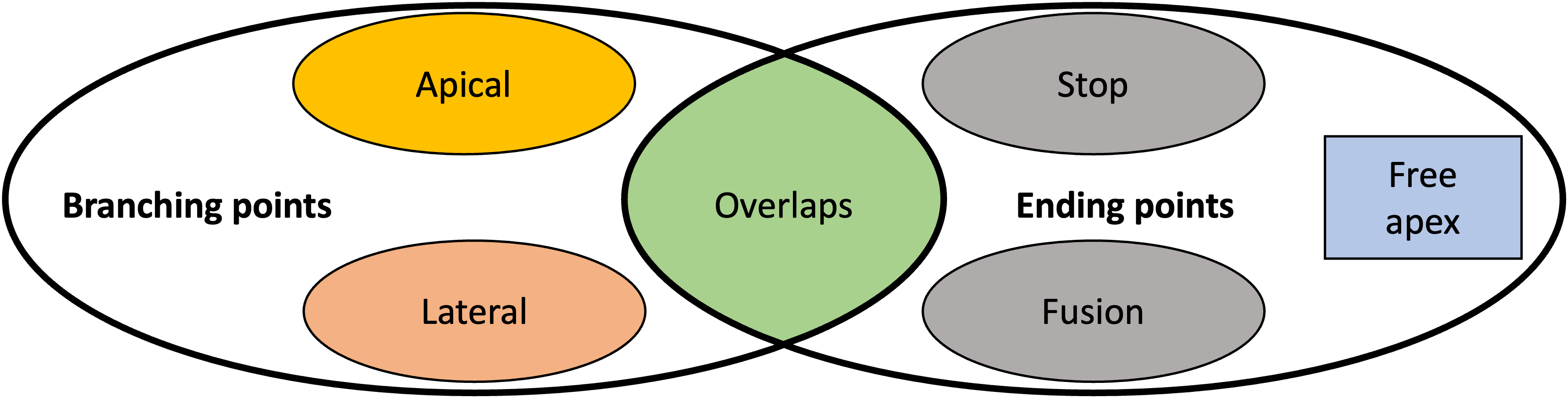}
    \caption{\textbf{Distinction between the degree of network vertices and their associated biological natures.} A degree-3 vertex may be due to a branching point which itself may be of two types (apical or lateral) but may also be the result of the meeting of two hyphae. In the case of a meeting, it is not possible to differentiate between anastomosis (fusion) and simple growth stop.}
    \label{fig:NodeClassification}
\end{figure}

As already discussed in the previous section, due to the image acquisition procedure, degree-3 vertices correspond to a collection of objects of different natures. Some are biologically relevant, others are not.
First correspond to branching points, \textit{ie} the departure of new hyphae, while others correspond to the meeting of two hyphae. 
Meeting point can be divided into two sub-categories. Anastomosis, \textit{ie} fusion of two hyphae, and encounters, if an apex arrives and stops growing close to a previously existing hypha. 
In the same way branching points can be divided into two sub-categories. Apical and lateral branching points, depending on the distance between the branch and the apex at the time of branching.
Overlaps are a population to be identified in order to correct spurious connections. Indeed, each overlap add two false degree-3 vertices, one appearing as a new branching point and the other one appearing as an encounters. This whole zoology of events is summarised in the figure \ref{fig:NodeClassification}.
Let-us now turn to the detection and classification of overlaps. The difficulty here comes from the particular topology of the crossing, that, in most cases, can not be reduced to simple crossing of two straight lines. As an example, the overlapping hypha may follow the overlapped hypha before resuming its own trajectory which, with a naive approach, would lead to the detection of two false branches. Conversely, an apex intercepting an hypha close to a previously existing branch would lead to the detection of a false overlap.  \\ 
We first consider all triplets of branches ($B_{carrier}$, $B_{in}$, $B_{out}$) where $B_{carrier}$ is the branch corresponding to the overlapped hypha, $B_{in/out}$ correspond to the incoming/outgoing part of the overlapping hypha. For the triplet to be a valid overlap, we checked a spatial and temporal conditions.
(i)-Connection condition: $B_{in}$ must end with a degree-3 vertex connected to $B_{carrier}$ and $B_{out}$ must start with a degree-3 vertex connected to $B_{carrier}$. We then derive the arc length $L_o$ corresponding to the distance between these two vertices. We set a threshold on $L_o$, typically a few hyphae diameters at which we consider that the triplet is no longer a potential overlap. \\
(ii)-Temporal condition: $B_{out}$ branch must have started its growth after the $B_{in}$ branch have arrived at the crossing. In fact, if we note $\Delta t_o$ the difference between the time of $B_{out}$'s growth start and the time of arrival of $B_{in}$, we must have $V \Delta t_o \approx L_o$ with $V$ the typical elongation rate of the network. \add{We find that the criterion $V\Delta t_o$ which must be in the range $ L_o \pm V\,T$ gives good results.}\\
Finally, of all the remaining triplets satisfying the above conditions, each branch can appear at most once as an incoming branch and at most once as an outgoing branch. If a branch appears several times in one of these roles, we keep only the triplet corresponding to the best alignment between the $B_{in}$ and $B_{out}$ branches.
When the overlap is confirmed, we disconnect the incoming and outgoing branches from the carrying branch and reconnect them together via a new node labeled negatively (to keep track of the overlap) situated at the mean position of the overlap. Therefore, the two degree-3 vertices become two degree-2 vertices along the carrying branch and the incoming and outgoing branches become the same branch.
\section*{Results}
\subsection*{Overlap detection performance}

\begin{figure}
    \centering{}
    \includegraphics[width=\linewidth]{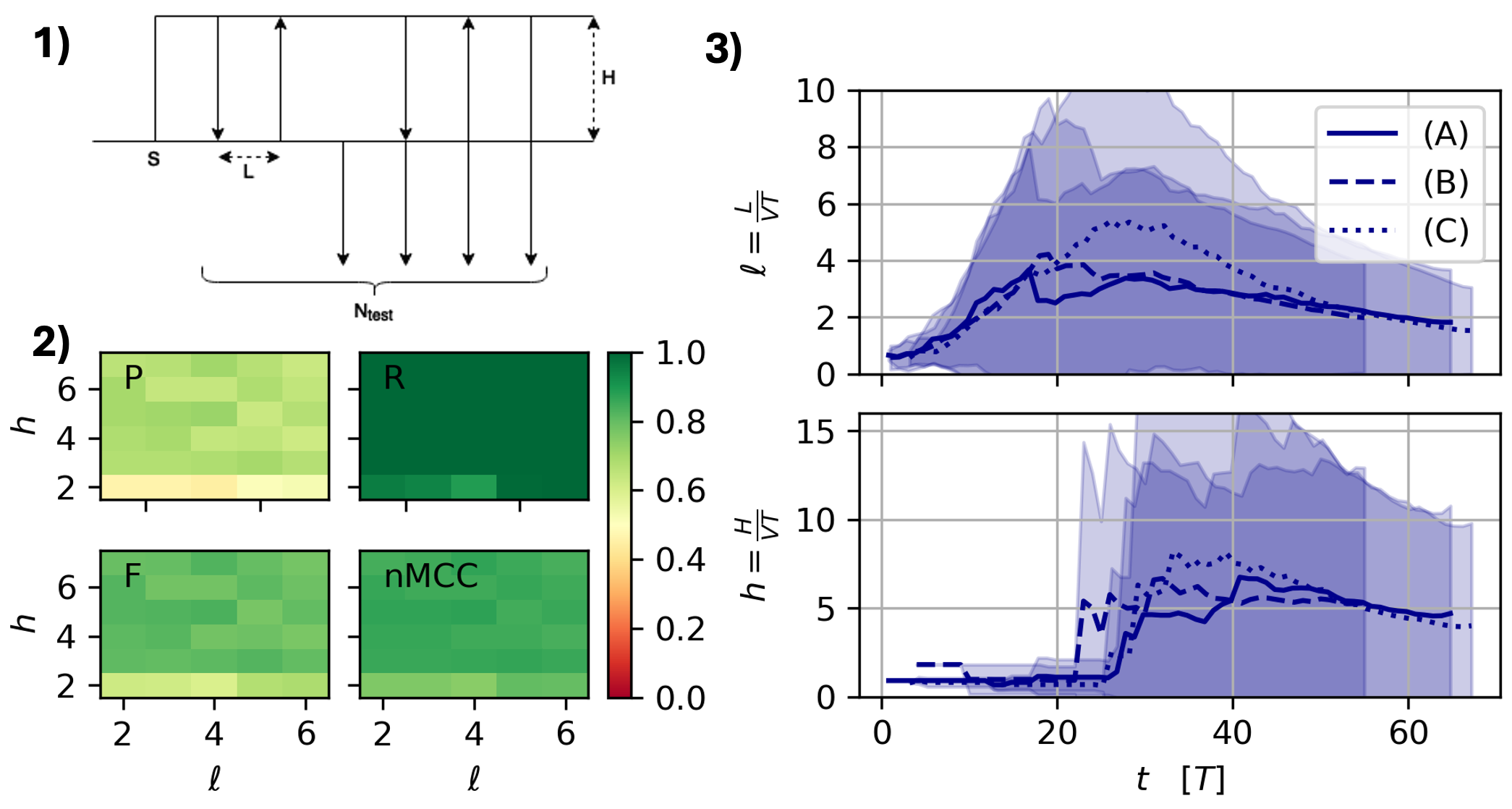}
    \caption{\add{\textbf{Overlap detection performance.}} \textbf{1)} Schematic diagram of the simulation reproducing the different hyphal encounter configurations. For each pair of $L$ and $H$ parameter values, ten simulations each generate $N_{test} = 100$ events randomly selected from the different encounter cases. \textbf{2)} Average precision $P$, recall $R$, F-score $F$ and normalized Matthew Correlation Coefficient  $nMCC$ of simulations as a function of $\ell$ and $h$ values. \textbf{3)} Measurements in the 3 networks (A), (B) and (C) of the $\ell=L/VT$ and $h=H/VT$ values encountered over time (mean + std).}
    \label{fig:results_simulation}
\end{figure}

\add{
	In order to estimate the performance of the overlap detection process, we build a simulation reproducing the different configurations observed. It consists in two parallel branches distant of length $H$. Branching process can occur along the two branches with length $L$ between them. $H$ and $L$ are normalized by the typical elongation length between two frames $V\,T$ where $V$ is the typical growth speed and $T$ the time between two frames. The two parameters of the simulation are therefore $h = \frac{H}{V\,T}$ and $\ell = \frac{L}{V\,T}$. The distance between the two branches is discretized, i.e. h is composed of integer values. For instance, $h=4$ means that 3 successive apex positions can be defined between the two main branches.
	The branching and encounter and overlaping possible configurations observed during the simulation are shown in figure \ref{fig:results_simulation}-1 and are described below.
	\begin{itemize}
		\item Branching from the top branch toward the bottom one and stop when the bottom one is reached.
		\item Branching from the bottom branch growing toward the top branch and stopping when the top branch is reached.
		\item Branching from the bottom branch and growing downward.
		\item Simultaneous branching from the top branch and the bottom branch, both going down at the same time.
		\item Double branching from the bottom branch, one growing toward the top branch and one growing downward.
		\item Branching from the top branch toward the bottom one then overlap and continue growing.
	\end{itemize}
	The order in which these processes occur along the main hyphae is chosen at random among this list.
} We set $N_{test}=100$ the number of events occurring along the main hyphae ; each $(h,\ell)$ configuration is reproduced ten times. 
To quantify the performance of the detection, we made use of usual binary classifications~\cite{Chicco2020Dec}. Precision $P = \frac{TP}{TP+FP}$, recall $R = \frac{TP}{TP+FN}$, F-score $F = 2\frac{PR}{P+R}$ and the normalised Matthews correlation coefficient $nMCC = (MCC+1)/2$ (with $MCC = \frac{TP\,TN-FP\,FN}{\sqrt{(TP+FP)(TP+FN)(TN+FP)(TN+FN)}}$) are calculated and reported in the figure~\ref{fig:results_simulation}-2 for various values of ($h$,$\ell$) around the typical range of values encountered in the experiments \add{(see figure \ref{fig:results_simulation}-3)}, with $TP$ the number of true positive overlap detection, $FP$ the number of false positive, $FN$ the number of false negative and $TN$ the number of true negative.
The main result of $P$ and $R$ shown  in figure \ref{fig:results_simulation} is that, except in a region where hyphae cannot be distinctly separated (both  $\ell$ and $h$ are small), detection quality is very good and relatively independent of $\ell$ and $h$.  More precisely, $R$ is found to be maximal, \textit{ie} FNs are very infrequently observed. $P$ is found in the range 0.7-0.9, which means a detection bias in favor of FP. Some branching are counted as overlaps.  
By combining the quantities $R$ and $P$, F-score summarizes the quality of the classification method used.
F-score ignores the count of True Negatives (TN). Because TN are supposed to be numerous, it is interesting to take into account all elements of the confusion matrix.  We find that $nMCC \sim 0.85$, a high confidence value, throughout the region of interest.
\add{The case where $h = 2$ gives a significantly worse prediction ($P\approx .5$), but with such a parameter, the twigs between the two main branches are too short. Only a single apex position can be used to determine the direction of growth of these twigs. This leads to overlaps being indistinguishable from the double branching process and therefore to a poor precision of the method.}
\subsection*{Distinction apical/lateral branching}
\begin{figure}
	\centering{}
	\includegraphics[width=\linewidth]{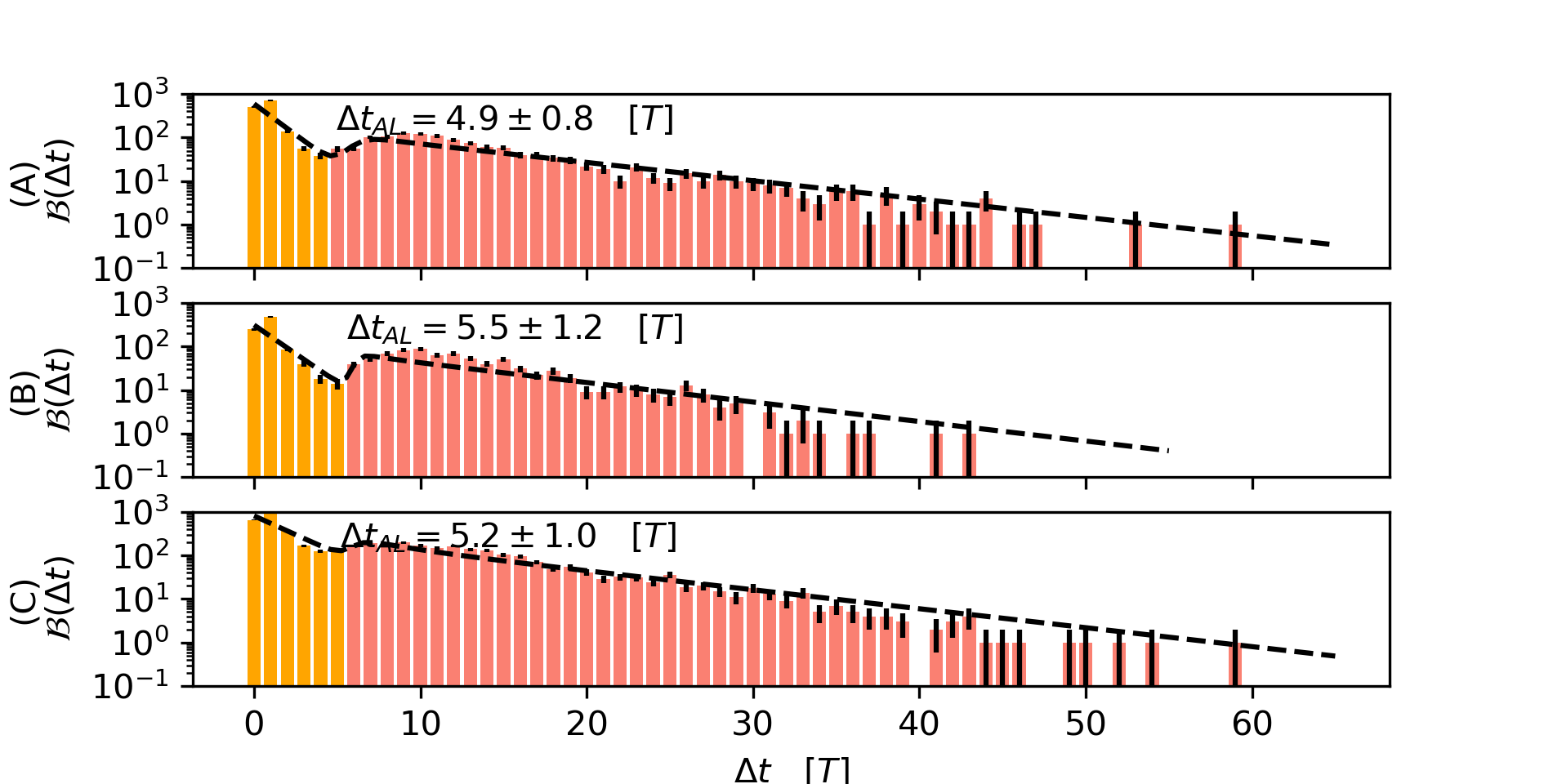}
	\caption{\textbf{Distribution of latency times before branching for the three experiments.} The black dashed line corresponds to the best fit based on equation \ref{eq:BranchingLatency}. Numerical values of the parameters of dashed lines are in table \ref{tab:Latency}. Values of $\Delta t_{AL}$ are calculated according to equation \ref{eq:ThresholdDeltat} and used to distinguish between apical branches (in orange) and lateral branches (in red).}
	\label{fig:Latency}
\end{figure}

Reduced to the most succinct list, the rules for network growth appear to be few. The first is polarized growth, the second is connection capacity. 
As shown in figure \ref{fig:NodeClassification} degree-3 vertices can be divided into two sub-populations, namely the lateral and apical branching points.
It seems odd that different processes are used to generate the same property. 
However in the literature the distinction between the two types is routinely made using  the location of their connection to the hypha. At the apex in the apical case, further away in the lateral case. This phenomenological distinction is linked to a lower growth rate for hyphae coming from lateral branching. In \cite{ledoux_prediction_2023} we proposed a classification criterion based on the distance apex/branching point, but it was not systematically evaluated on the whole thallus, making it possible for a bias to exist. In this section, building our analysis on the complete thallus we propose a definitive criterion to clearly distinguish between apical and lateral branching.
To remove the ambiguity of the distance to the apex during encounters, we look instead at the latency time $\Delta t$ between the first passage of the mother branch and the time the new branch starts to grow as shown in figure \ref{fig:Latency}.

We observe a bi-modal distribution where we assume that the first mode corresponds to the apical branching process while the second is the result of lateral branching process. In fact, by looking at a single branch at a given time we can easily convert the latency before branching to a length from the tip assuming a constant elongation rate. Therefore the shortest latency are closest to the tip and should correspond to apical (or sub-apical) branching points. 
We can express $\mathcal{B}_{t_i}^{t_f}(\Delta t)$ the number of branching observe between the instant $t_i$ and $t_f$ corresponding at a latency of $\Delta t$ as the sum of the corresponding apical branching points $\mathcal{A}_{t_i}^{t_f}(\Delta t)$ and lateral branching points $\mathcal{L}_{t_i}^{t_f}(\Delta t)$. 
Assuming that apical branches are distributed according to an exponential law (parameter $\omega_A$) with respect to the apex and that lateral branches are uniformly distributed along the hypha except for an exclusion zone near the apex (modelled by a sigmoid centred at $m$ and with slope $s$) one can show that $\mathcal{B}_{t_i}^{t_f}(\Delta t)$ is given by (see appendix for more details):
\begin{equation}
	\begin{split}
\mathcal{B}_{t_i}^{t_f}(\Delta t) &=\tau_A \frac{N(t_f)-N(t_i)}{\nu \log{2}} e^{-\omega_A \Delta t}\frac{1-e^{-\omega_A T}}{1-e^{-\omega_A(t_f-t_i+T)}}\\
		 &+\frac{\tau_L}{1+e^{-(\Delta t-m)/s}} (1-2^{-\lambda T})\frac{L(t_f-\Delta t) - L(t_i)}{\lambda \log{2}}
	\end{split}
    \label{eq:BranchingLatency}
\end{equation}
With $(L_0, \lambda),(N_0,\nu)$ the parameters of the exponential fit of  respectively the total length of the network $L$ and the total number $N$ of branching points.
We can then define $\Delta t_{AL}$, the threshold below which the branch is considered apical as the ``starting point" of the lateral branching sigmoid:
\begin{equation}
    \Delta t_{AL} = m - 2 s
    \label{eq:ThresholdDeltat}
\end{equation}
We find that \add{$\Delta t_{AL} = 5.2\pm0.6 [T] $ which converts to} $\Delta t_{AL} = 94\pm11 [min]$. 

\begin{table}[]
    \centering
    \begin{tabular}{|c||c|c|c||c|}
    \hline
    Experiment & (A) & (B) & (C) & Mean\\\hline\hline
    $N_0$ & $3.19\pm0.09$ & $3.7\pm0.2$ & $1.9\pm0.1$ & Not applicable\\\hline
    $\nu [h^{-1}]$ & $0.506\pm0.003$ & $0.540\pm0.004$ &$0.570\pm0.006$ & $0.539\pm0.006$\\\hline
    $L_0 [mm]$ & $1.09\pm0.05$ &  $1.52\pm0.08$ & $1.00\pm0.04$ & Not applicable \\\hline
    $\lambda [h^{-1}]$ & $0.476\pm0.004$ & $0.501\pm0.005$ & $0.508\pm0.004$ & $0.495\pm0.003$ \\\hline\hline
    $\tau_A [h^{-1}]$ & $0.15\pm0.02$& $0.14\pm0.02$ & $0.18\pm0.02$ & $0.16\pm0.01$\\\hline
    $\omega_A [h^{-1}]$ & $2.2\pm0.2$ & $2.0\pm0.3$ & $1.3\pm0.2$ & $1.8\pm0.2$\\\hline
    $\tau_L [h^{-1}.mm^{-1}]$ & $0.96\pm0.09$ & $0.9\pm0.1$ & $1.1\pm0.1$ & $0.97\pm0.06$\\\hline
    $m [min]$ & $107\pm7$ & $109\pm6$ & $108\pm9$ &$108\pm4$\\\hline
    $s [min]$ & $9\pm7$ & $4\pm8$ & $7\pm8$ & $7\pm5$\\\hline\hline
    $\Delta t_{AL} [min]$ & $88\pm15$ & $100\pm22$ & $93\pm18$ & $94\pm11$\\\hline
    \end{tabular}
    \caption{Numerical values from the fit of latency from figure \ref{fig:Latency} according to equation \ref{eq:BranchingLatency}. $\Delta t_{AL}$ is calculated from $m$ and $s$ as in \ref{eq:ThresholdDeltat}.}
    \label{tab:Latency}
\end{table}

\subsection*{Distinctions between apical and lateral branches}
\begin{figure}
    \centering
    \includegraphics[width=\linewidth]{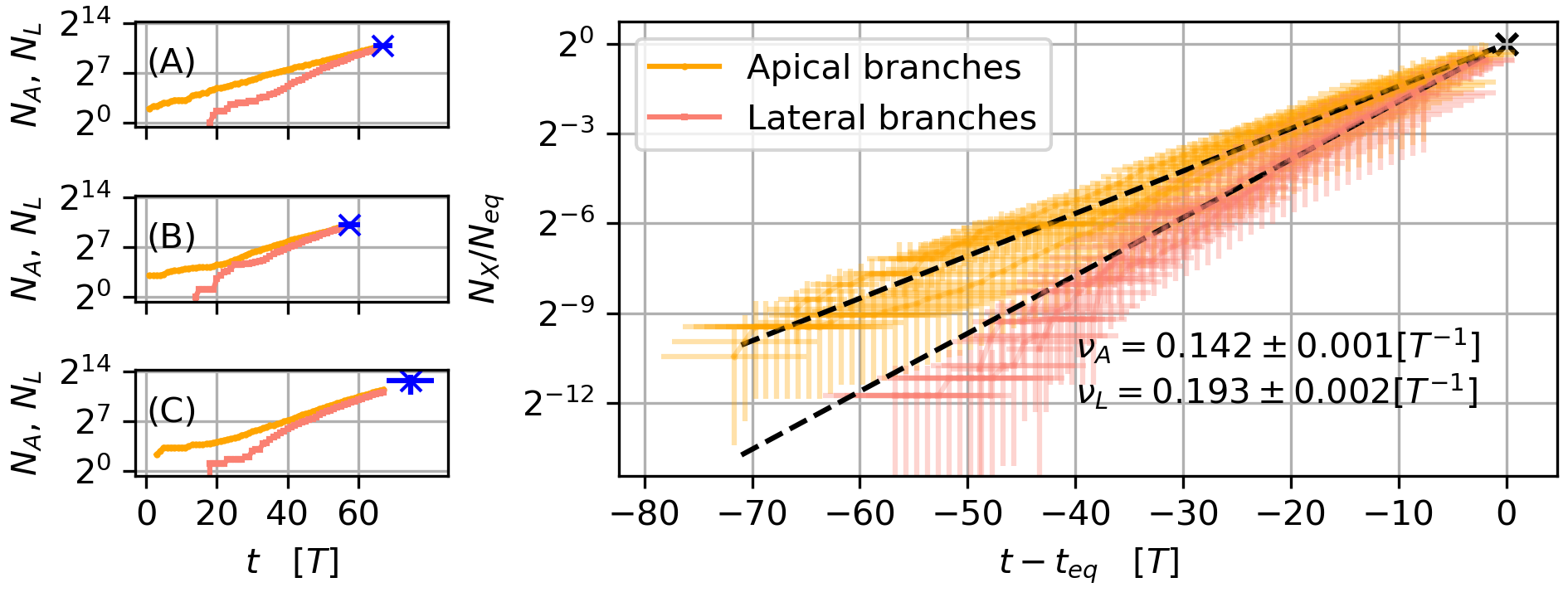}
    \caption{\add{\textbf{Temporal distinction between apical and lateral branches.}} Left) counts of apical \add{($N_A$)} and lateral \add{($N_L$)} branch over time for the three experiments.  Blue marker correspond to the point $(t_{eq},N_{eq})$ for each experiment. Right) normalized count of apical and lateral branch over unified time for the three experiments. Dashed line correspond to mean exponential fits for apical (in orange) and lateral (in red).}
    \label{fig:TemporalDistinction}
\end{figure}

\begin{figure}
	\centering
	\includegraphics[width=\linewidth]{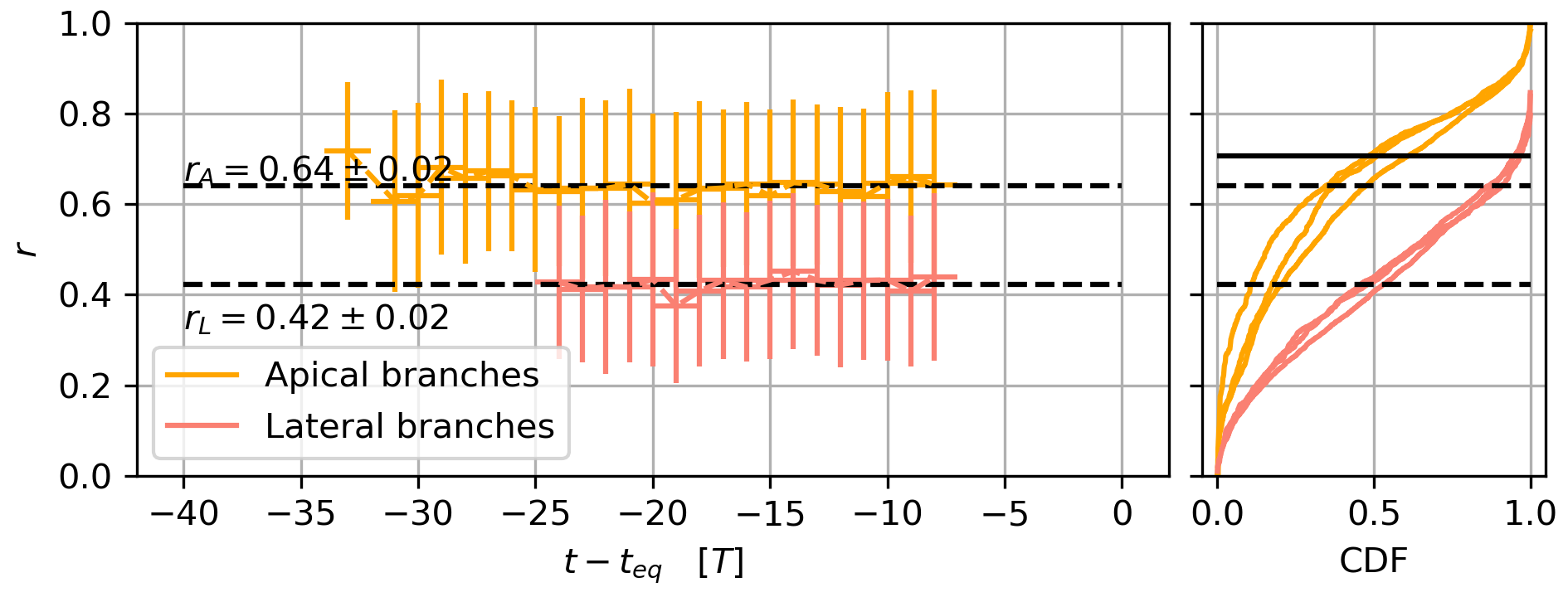}
	\caption{\add{\textbf{Spatial distinction between apical and lateral branches.}}Left) Ratio of the euclidean distance to the spore over the maximal radius at the moment of branching for the apicals in orange and the laterals in red (mean $\pm$ standard deviation) as a function of time. Right) Cumulative density estimate of this distributions. Dashed line corresponds to $r_A$ mean value for apical branchings and $r_L$ mean value for lateral branching. Solid black line correspond to the theoretical division where the outer ring has the same area as the inner disk at $r = \frac{\sqrt{2}}{2}$.}
	\label{fig:SpatialDistinction}
\end{figure}

\begin{figure}
	\centering
	\includegraphics[width=\linewidth]{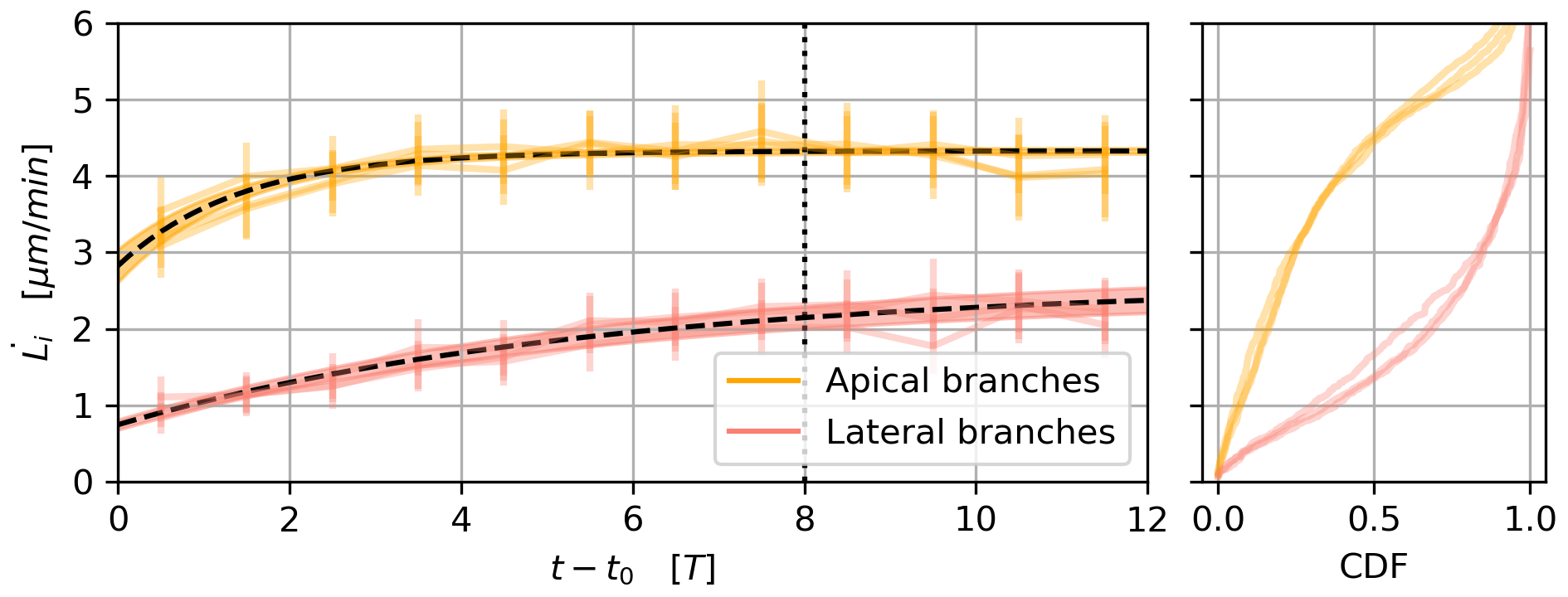}
	\caption{\add{\textbf{Dynamical distinction between apical and lateral branches.}} Left) Mean elongation $\dot{L_i}$ as a function of time since the branch start $t-t_0$ for both apical (in orange) and lateral (in red) branches. Dashed line correspond to the fit according to Equation \ref{eq:Elongation}. \add{Vertical dotted line correspond to the start of stationary domain ($t-t_0\geq 8T$).}
	Right) Cumulative density estimate of elongations in the stationnary domain ($t-t_0\geq 8T$).}
	\label{fig:DynamicalDistinction}
\end{figure}

\paragraph{Temporal distinction:}
This distinction allows us to monitor the respective number of apical and lateral branches over time as we can see in figure \ref{fig:TemporalDistinction}. Both follow an exponential growth of the form:
\begin{equation}
    N_X = N_X^0\base^{\nu_X \,t}
    \label{eq:ApicalLateralCount}
\end{equation}
At first, branches come only from apical branching. However, the effective rate of lateral branching is greater than that of apical branching. This leads to the point where both kind of branches are present in equal number.
We can extract this time $t_{eq}$ when $N_A(t_{eq}) = N_L(t_{eq}) = N_{eq}$: 
\begin{equation}
    t_{eq} = \frac{\log_2 \left(N_L^0/N_A^0\right)}{\nu_A-\nu_L}
    \label{eq:t_eq}
\end{equation}
in the following we use this time as a reference to synchronize the three experiments (see figure~\ref{fig:TemporalDistinction}).
\paragraph{Spatial distinction:}
Another metric interesting to look at is the respective position of such branching points. 
For this we consider the distributions of the ratio of the euclidean distance to the spore over the maximal radius of the thallus at the moment of branching for the two kinds of branching (see figure \ref{fig:SpatialDistinction}). 
In both apical and lateral case, this ratio appears to stabilize around a constant value over time although the maximum radius grows linearly with time. The difference in the value of this constant, $r_A = 0.65\pm 0.03$ in the apical case and  $r_L = 0.42 \pm 0.02$ in the lateral case, reflects a difference in the spatial occupation of these branches. Lateral branches are mainly found inside the thallus, while apical branches are found in the outer ring. Cumulative density estimate in figure \ref{fig:SpatialDistinction} show that half the apical branches appear above  $r\approx 0.66$ while almost all lateral branches appear below this value.
\paragraph{Dynamical distinction:}
Another distinction between apical and lateral branches is their elongation rate.  In Figure~\ref{fig:DynamicalDistinction}, we have plotted the mean elongation $\dot{L_i}$ for branches emerging from an apical branching point or a lateral one according to the time since the branch start growing (noted $t_0$ in the figure). We considered only branches growing over at least 12 frames, that corresponds to approximatively $860\,\mu m$ long branch for apical ones and $625\,\mu m$ for lateral ones. From respectively experiments $A$, $B$ and $C$ we extracted  96, 82 and 98 apical branches and 50, 51 and 53 lateral branches.
 We observe that in both cases the mean elongation start by an acceleration phase at an initial value $\dot{L_i^0}\neq 0$ before reaching a steady-state at rate $\dot{L_i^f}$. We fitted the result according to the following empirical law:
\begin{equation}
	\dot{L_i}(t) = (\dot{L_i^f}-\dot{L_i^0})\left(1-e^{-\gamma t}\right) + \dot{L_i^0}
	\label{eq:Elongation}
\end{equation} 
For apical branches we find $\dot{L_i^0} = 2.8\pm 0.2 \mu m/min$, $\dot{L_i^f} = 4.33\pm 0.03 \mu m/min$ and $\gamma = 2.3\pm 0.5 h^{-1}$ while for lateral branches we find $\dot{L_i^0} = 0.74 \pm 0.06 \mu m/min$, $\dot{L_i^f} = 2.6\pm 0.2 \mu m/min$ and $\gamma = 0.6\pm 0.1 h^{-1}$.

\subsection*{Count of the number of free apexes, branching points, encounters and overlaps.}

\begin{figure}
    \centering
    \includegraphics[width=\linewidth]{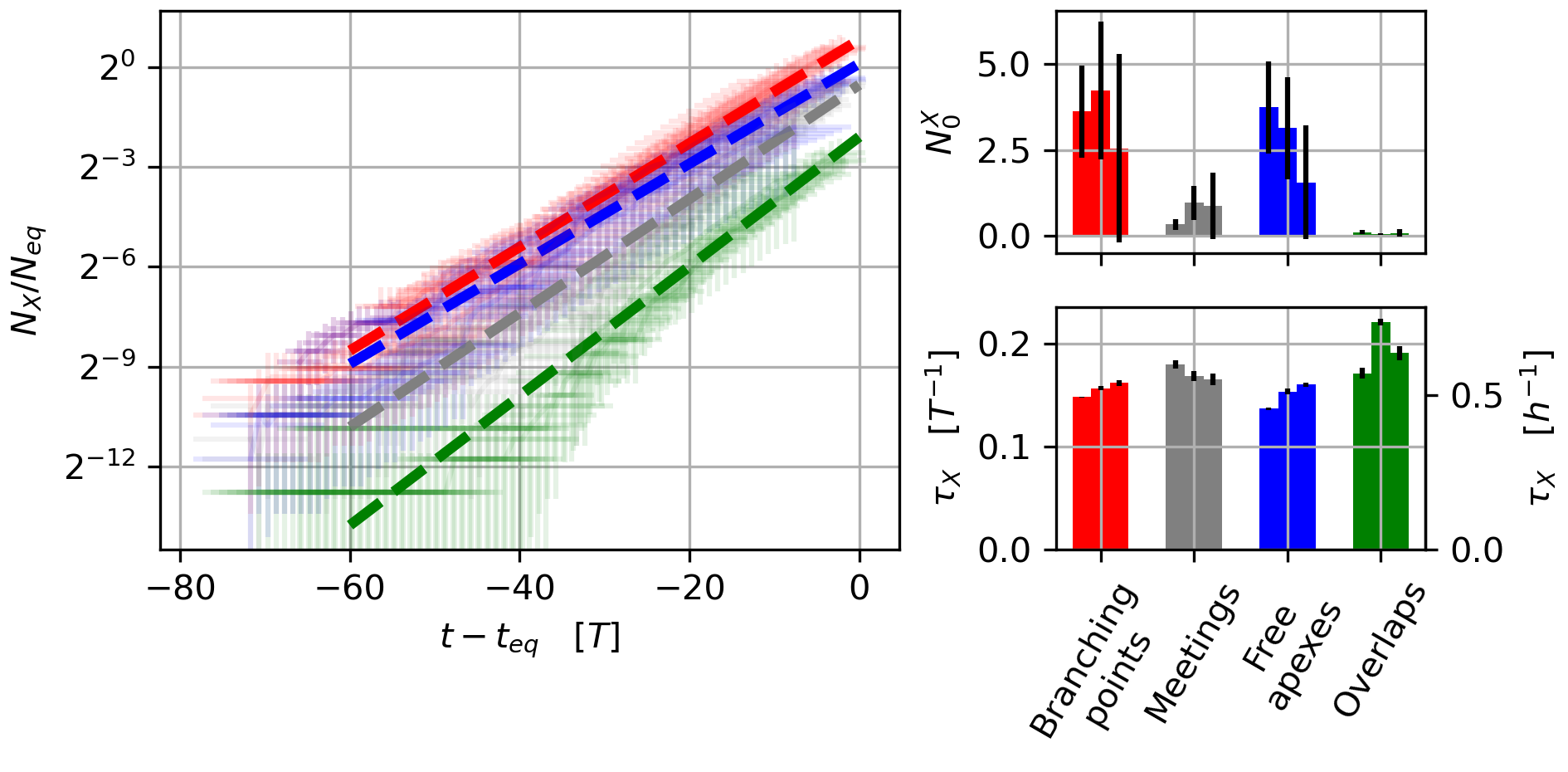}
    \caption{\add{\textbf{Direct counting of the number of biological objects in the network.}} \textbf{1)} Counts of branching points (in red), free apexes (in blue), encounters (in grey) and overlaps (in green) with respect to time normalized for each experiment (A), (B) and (C). Dashed lines correspond to mean fits for each counts of the form $N_X = N_0^X 2^{\tau_X t}$.
    \add{\textbf{2)} Numerical values of the fits parameters $N_0^X$ (top) and $\tau_X$ (bottom)}}
    \label{fig:Counts}
\end{figure}

Dealing with a network cleaned of overlaps, with branches rebuilt in their entirety, we are now able to distinguish the different biological origins of degree-3 vertices.  Direct counting of the number of biological objects in the network, branching point (lateral and apical), free apexes and encounters are shown in figure~\ref{fig:Counts}. Encounters combine anastomosis and cessation of apex growth in the immediate vicinity of a hypha. To complete the description with non biological structures, the number of overlaps was added into the figure. 
It was reported in \cite{dikec_hyphal_2020} an exponential growth over time of the total length of the network. In this work,  we observe the same exponential behavior for the numeration of branching points, free apexes, encounters and overlaps. We adjust these exponential growths in figure \ref{fig:Counts} according to the same expression as for the growth in number of apical and lateral branches described in equation \ref{eq:ApicalLateralCount} with $X$ corresponding this time to branching points, free apexes, encounters or overlaps.
As expected, at the beginning of the growth, with a very simple network, only branching points and free apexes are observed.  With the increase in density, 
encounters and overlaps become more frequent. At the end of the experiment, about half of the branches no longer end in free apexes, but in an encounter.
Interestingly,  overlaps and encounters  growth rates are approximately similar, which suggests that the proportion of encounters that become overlaps does not depend on the stage of growth.  This proportion (Number of overlaps divided by the sum of the number of overlaps and encounters) is calculated to be in average of $19\pm1\%$.
We also note that the variability between experiments is greater than the variability between the different objects studied here (branches, free apexes, encounters and overlaps).
\section*{Discussion}
In this work, we propose to exploit images of the growth of the filamentous fungus thallus \Podo{}, model of a growing and branching network.
These images were obtained using a simple microscopic device, \textit{i.e.} exploiting the possibilities offered by recent developments in image capture systems and the automation of spatial and temporal monitoring. 
Despite its advantages, fluorescence lighting has not been chosen. Our experimental system has the advantage of being sufficiently efficient in white light. This is an advantage insofar as, whenever possible, the use of white light illumination is preferable to fluorescence illumination, as it alleviates the experimental constraints linked to the fluorescent illumination device, the constraints linked to reconstruction (for instance alignment can be made tricky), and the effects of tissue ageing.
Growing on a cellophane sheet enables two-dimensional guidance of the thallus, the simplicity of which allows high sampling frequency and easy reconstruction compared with three-dimensional techniques. 
The result is a collection of independent images that can be represented by a single graph, made up of interconnected nodes of degree-3 or less. Individual labeling of each object in the network is made problematic by the lack of articulation between images. In particular, the overlaps introduce an over-numeration of nodes of degree-3.
Thanks to the tracking of apex trajectories through individual branch identification, we can identify overlaps as a combination of a branch ending in an encounter and a new branch synchronized in time and space along the same carrier branch. 
Comparison with a simulation has enabled us to estimate the good efficiency of our method in the range of parameter values encountered within experiments through different metrics such as the F-score and the Matthew Correlation Coefficient.
Thanks to this reconstruction, we can identify the different branches of the network, allowing us to track any apex while retaining global information on network connectivity. 
To automatically determine the nature of this branches (apical or lateral), we propose a latency time criterion between the first passage of a hypha and the moment of branching. Below the threshold value $\Delta t_{AL}$, whose numerical value is obtained by adjusting latency times at the scale of the complete thallus, the branching is considered apical (and inversely lateral).
At the start of the exponential growth phase, the network is composed solely of branches from an apical branch. However, as the mechanics are different in the two mechanisms (proportional to the number of apexes for apical branches and proportional to the total length of the network for lateral branches), the effective growth rate of lateral branches is higher than that of apical branches. 
This distinction removes the ambiguity surrounding the initial latent phase due to germination. This phase of random duration precedes the network's exponential growth phase. This makes it difficult to compare one experiment with another. We then define the instant $t_{eq}$ when the two sub-populations (apical and lateral branches) are in the same number, \textit{i.e.} $N_A(t_{eq}) = N_L(t_{eq})$. The latter can be used as a reference to compare growth times between different experiments.
As reported in \cite{ledoux_prediction_2023}, the density of the network depends directly on the branching activity of the apical and lateral types.  The former follows an exponentially decreasing law from the apex, while the latter shows a continuous branching rate per unit length. The time lag observed between the two processes spontaneously generates a density minimum. We show here that the time lag is not the only factor involved, but is also coupled with a spatial difference in branch organization and a difference in branch growth dynamics.
The thallus is an inhomogeneous object. The biological nature of the branches depends on their location relative to the origin (the spore) of the site observed, and the time between it and the passage of the growth front. The biological nature corresponds to lateral and apical branches. These two sub-populations grow in the same network, but with different branching mechanisms and elongation speeds. 
These behavior discrepancies are probably linked to various biological processes at cellular level, still largely unknown, and governing both types of branching. Then, while apical branching usually involves a significant disturbance in the growth rate and morphology of the parental hyphal tip, with a temporary disappearance of the Spitzenkörper, lateral branching occurs without any detectable alterations in the growth or Spitzenkörper behaviour of the parental hypha \cite{riquelme_architecture_2011}.
Branches from apical branches are faster, earlier and located in the outer crown of the thallus. These characteristics, in addition to the asymmetrical branching angles \cite{ledoux_prediction_2022}, make them ideal for exploration.
Lateral branches are slower, later and tend to be located in the center of the thallus. These features, in addition to the typical 90-degree branching angles also reported in \cite{ledoux_prediction_2022}, make them densification hyphae.
It should be noted here that these results are in line with the functional description of the fungal thallus, reported several years ago. Then, \cite{lew_mass_2005} described the fungal network as consisting of apical cells (or leading hyphae) which are the first cells to invade new territory and are generally engaged in nutrient acquisition and sensing of the local environment, whereas behind the colony edge, sub-apical cells generate new hyphae by lateral branching. An earlier description of such fungal network states that hyphal tips at the biomass edge are those associated with exploration while hyphal tips behind the biomass front are most associated with resource exploitation \cite{boswell_functional_2002}.
It is important to note that the distinction proposed in this work is based on a statistical approach. 
In the absence of a clear distinction between apical and lateral differences, the proposed studies were based on spatial location, with apical branches assumed to be favorably furthest from the spore. 
We justify this choice by noting the predominance of apical branches in the outer ring. We can be more precise and defined the outer ring according to the distance between the spore and the furthest apex. Beyond $r = \sqrt{2}/2$ there are statistically no more lateral branches, which corresponds to a division of the thallus into two zones, the outer ring and inner disk, of equal areas.
Conversely, the type of branches present in the densest regions is statistically composed of lateral branches. To our knowledge, their specific study has never been undertaken.
The reconstruction is based on the characteristics of the fungal thallus studied. Some of these can be lifted, enabling this work to be generalized to any growing network in two dimensions, biological or not, e.g. on cities or fractures. The origin of the network is assumed to be unique. This is necessary in the reconstruction of complex branches, whose causality cannot be deduced from observation, especially with very short segments. The multiple-source assumption is not allowed in the case presented here. However, this configuration can be found with growth ignition based on multiple ascospores or from fragmented mycelia. In this case, one way to overcome this difficulty could be to recognize them individually before processing.
The filaments making up the network are considered identical, \textit{i.e.} their diameter is not taken into account. This information could be easily retrieved, enabling the study of networks that reorganize themselves such as the one described in the slime mouls Physarum polycephalum \cite{reid_thoughts_2023} and adapt the flows passing through them by modifying their characteristics.
Another characteristic of the blob is not taken into account in this work. A network can evolve by abandoning low-use connections, which can lead to necrosis. This work is based on a  reference image containing the entire network. Insofar as the growth of \Podo~ is not accompanied by retraction, a late image is compatible with this work. A reorganization of the network generated by an abandonment of connections, could be taken into account if the network were found to evolve between different well-defined states of maximum complexity. An extraction of the images corresponding to these states would enable the intermediate states to be reconstructed. Similarly, the network is expected to be fixed in the laboratory reference frame. If a node or connection were to slip, it could not be recovered.
\section*{Conclusion}
Overall, the reconstruction of the dynamics of this network allows us to follow each apex over time (despite overlaps) and to identify each hypha individually, with the possibility of automatically categorizing the nature of the branching (apical or lateral) at its origin. This labelling makes it possible to automate several measurements that can then be applied rapidly to the entire network, such as measuring branching angles or elongation rates. The differences in dynamics and temporality according to the type of hyphal branching also highlight the different roles of hyphae between exploration and exploitation of the environment.

\section*{Author Contributions}
F.C.L. and E.H. initiated the research. T.C. wrote the data extraction programmes and analyzed the data. All authors discussed the results and wrote the article.

\section*{Acknowledgments}
The authors would like to thank Clara Ledoux for conducting the experiments, Frederic Filaine for the experimental set-up, Christophe Lalanne for his advice on statistical analyses, Aurélien Renault for the technical assistance. and the NEMATIC group for fruitful discussions. Thibault Chassereau is supported by a PhD scholarship from ANR-21-CE40010-01.

\section*{Declaration of interests}
The authors declare no competing interests.

\section*{Appendix}
\subsection*{Distribution of latency before branching}
We aim to derive an expression for the observable $\mathcal{B}_{t_i}^{t_f}(\Delta t)$, the number of branchings occuring during the period $t \in [i \, T;f \, T[$ and appearing after a latency time  $\Delta t \in [j \, T ; (j+1) \, T[ $ , where $1/T$ is the frame rate and $i,f, j$ are integers. The time origin is defined locally for each network node by the growth of the hypha at that point. Technically, the passage of an apex is detected. $\mathcal{B}_{t_i}^{t_f}(\Delta t)$ can be expressed as the sum of the corresponding apical branching $\mathcal{A}_{t_i}^{t_f}(\Delta t)$ and lateral branching $\mathcal{L}_{t_i}^{t_f}(\Delta t)$:
\begin{align}
    \mathcal{B}_{t_i}^{t_f}(\Delta t) &= \mathcal{A}_{t_i}^{t_f}(\Delta t)+\mathcal{L}_{t_i}^{t_f}(\Delta t)
    \label{eq:B}
\end{align}

\begin{figure}[!h]
	\centering
	\includegraphics[width=0.75\linewidth]{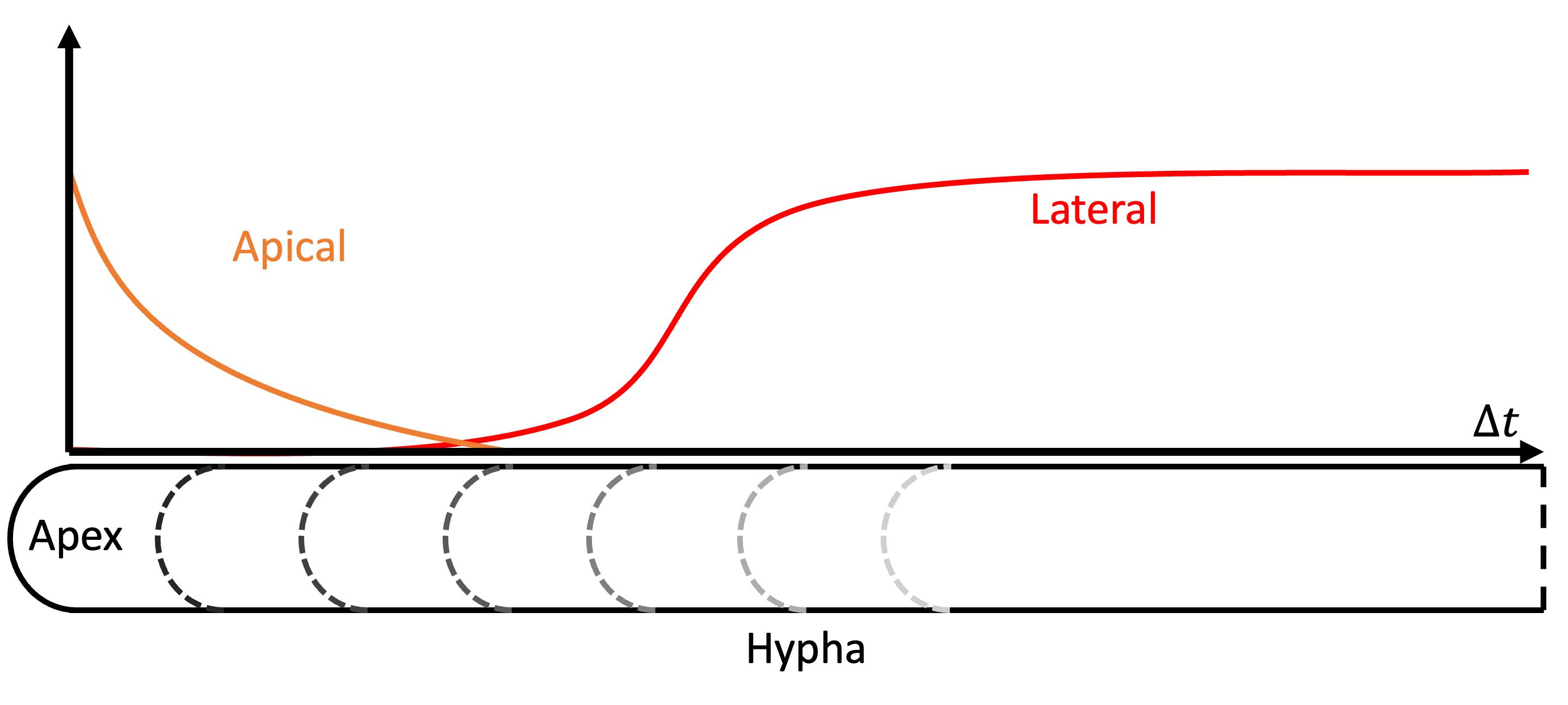}
	\caption{Diagram illustrating the distribution of apical (in orange) and latera (in red) branches along a single hypha.}
	\label{fig:repartitionBranchements}
\end{figure}
\paragraph{Apical part}
In this section we will build $\mathcal{A}_{t_i}^{t_f}(\Delta t)$ the apical contribution of $\mathcal{B}_{t_i}^{t_f}(\Delta t)$. Introducing $A(t_i,t_f)$ as the total number of apical branchings appearing between $t_i$ and $t_f$ and $\phi_A(\Delta t)$ as the probability for an apical branching to occur after $\Delta t$, we then write:
\[
 \mathcal{A}_{t_i}^{t_f}(\Delta t) = A(t_i,t_f) \, \phi_A(\Delta t)
\]
At each time step, the number of apical branches appearing in the network is proportional to the number of apexes present in the network, which corresponds to the number of branches $N$ (from apical or lateral branching). With $\tau_A$ the proportionality coefficient, we write $A(t_i,t_f)$ as 
\[
A(t_i,t_f) = \tau_A \, \int_{t_i}^{t_f}N(t)\, \text{d} t =\tau_A \, \frac{N(t_f)-N(t_i)}{\nu \log{2}}
\]
with $N(t) = N_0 \, 2^{\nu t}$. 
Following \cite{ledoux_prediction_2023} we expect apical branching  on a single hypha to be exponentially distributed from the tip, with a parameter $\omega_A$.  Following we can derive $\phi_A(\Delta t)$ as
\[
 \phi_A(\Delta t) = e^{-\omega_A \Delta t} \, \frac{1-e^{-\omega_A \, T}}{1-e^{-\omega_A\,(t_f-t_i+T)}}
 \]
which is normalized over all latencies, \textit{i.e.} such that $\sum_{\Delta t = 0}^{t_f-t_i} \phi_A(\Delta t) =1$.

\paragraph{Lateral part}
We will now focus on $\mathcal{L}_{t_i}^{t_f}(\Delta t)$. Along a single hypha, we expect lateral branching to be distributed uniformly along the hypha except in the dominance apical region near the tip (see Figure \ref{fig:repartitionBranchements}). We modelized this property by modulating the lateral branching rate $\tau_L$ according to a sigmoid of parameter $m$ and $s$ depending on the latency $\Delta t$. Unlike the apical branchings, we assume that the number of lateral branchings appearing at time $t$ is proportional to the total length of the network $L(t)=L_0\,2^{\lambda\,t}$.
By noticing $\frac{\tau_L }{1+e^{-(\Delta t-m)/s}}$ does not depend on $t$, we get:
\begin{align*}
    \mathcal{L}_{t_i}^{t_f}(\Delta t) 
        &= \int_{t_i+\Delta t}^{t_f} \, \text{d}t 
            \, \frac{\tau_L }{1+e^{-(\Delta t-m)/s}} 
            \, \left(L(t-\Delta t) - L(t-\Delta t - T)\right)\\
        &=\frac{\tau_L }{1+e^{-(\Delta t-m)/s}} 
            \, (1-2^{-\lambda T})
            \, \frac{L(t_f-\Delta t)-L(t_i)}{\lambda \log{2}}
\end{align*}
Finaly equation \ref{eq:B} can be rewriten to obtain equation~(4) of the main text:
\begin{align}
    \mathcal{B}_{t_i}^{t_f}(\Delta t) 
        &= \tau_A \frac{N(t_f)-N(t_i)}{\nu \log{2}} e^{-\omega_A \Delta t}\frac{1-e^{-\omega_A T}}{1-e^{-\omega_A(t_f-t_i+T)}} \nonumber\\
        & +\frac{\tau_L}{1+e^{-(\Delta t-m)/s}} (1-2^{-\lambda T})\frac{L(t_f-\Delta t) - L(t_i)}{\lambda \log{2}}
        \label{eq:B2}
\end{align}
$\nu$, $N_0$, $\lambda$ and $L_0$ can be determined from the fits of respectively $N$ and $L$ as a function of time $t$. 
Following, fit $\mathcal{B}_{t_i}^{t_f}(\Delta t)$ (see figure~(6) of the main text) allowed us to extract $\omega_A$, $\tau_A$, $\tau_L$, $m$ and $s$. The corresponding experimental values are shown in table \ref{tab:Latency}.
\subsection*{Pseudo-code algorithms}\label{annexe:PseudoCodes}
\begin{algorithm}[!h]
    \caption{Global pipeline.\\
        A pictorial summary of this algorithm pseudo-code is proposed figure \ref{fig:recap}}
    \label{alg:GlobalProcedure}
    \begin{algorithmic}[]
        \Require Stacks of growth images, named \textit{images} hereafter.
        \State  \textbf{MANUAL BINARISATION}(first image) $\rightarrow$ first binarised image  
        \State  \textbf{MANUAL BINARISATION}(final image) $\rightarrow$ final binarised image
        \State  \textbf{VECTORISATION}(final binarised image) $\rightarrow$ spatial graph 
        \State \textbf{DATING}(spatial graph, images, first bin. image, last bin. image) $\rightarrow$ dynamical graph 
        \State  \textbf{TWIGS ORIENTATION}(dynamical graph) $\rightarrow$ oriented twigs graph 
        \State  \textbf{BRANCH IDENTIFICATION}(oriented twigs graph) 
        $\rightarrow$ branched graph 
        \State  \textbf{OVERLAPS CORRECTION}(branched graph) 
        $\rightarrow$ final graph
    \end{algorithmic}
\end{algorithm}

\begin{algorithm}[!h]
	\caption{Function: DATING\\ 
        Give each node of spatial\_graph a time coordinate. In a second phase, ensure that the graph is connected at all times.}\label{alg:Dating}
	\begin{algorithmic}[]
    		\Require spatial\_graph, images, first binarised image, last binarised image \\

    		\For{each pixel lit in the last binarised image }\Comment{Pixel dating}
    				\State find the pixel activation time by fitting the time series of light intensity using the step function $S_{t_0}$.
    		\EndFor
    		\For{ each node in spatial\_graph} \Comment{First estimation of $t_0$}
                \State find, in a circle of radius $R$ (here $R = 3$ pixels) centered on the node, the time $t_0$ when the first pixel lit
    		\EndFor
    		\For{each node $n$ in spatial\_graph} \Comment{First correction}
                \State extract $t_0$ of each adjacent nodes. $t_{min}$ is the minimum of this collection.
    			\If{$t_0$ of node $n$ is strictly less than $ t_{min}$} 
                    \State $t_0$ of node $n$ is set to $ t_{min}$
    			\EndIf
    		\EndFor 
    		\\\Comment{Neighbor to neighbor correction}
            \State Build $g_0$, the subgraph from spatial\_graph compatible with the mask made by the first binarised image (\textit{i.e.} the ascospore)
    		\For{each timestep $t$}
    			\State Initialise the subgraph $g_t$ by taking the subgraph corresponding to the previous timestep.
    			\Repeat
    			\State Add to the subgraph $g_t$ every nodes that are not in $g_t$, adjacent to at least one node of $g_t$ and whose estimation of $t_0$ is less than or equal to $t$.
    			\State Set the $t_0$ estimation of all added nodes to $t$
    			\Until{no more nodes are added.}
    		\EndFor
    	\end{algorithmic}
\end{algorithm}

\begin{algorithm}[!h]
	\caption{Function: TWIGS\_ORIENTATION\\
	Direct every twigs of dynamical\_graph according to the growth orientation}\label{alg:TwigsOrientation}
	\begin{algorithmic}[]
		\Require dynamical\_graph\\
		\State Select the position of the source as the barycenter of the initial graph (subgraph of dynamical\_graph where we keep only the nodes existing at time $0$) and fix the orientation of the three twigs emerging from it.
		\State Orient all the terminal twigs (ending with a degree 1 node) and propagate this orientation recursively knowing that only one source, located at the ascospore is possible.
		\For{each twig in dynamical\_graph}
			\State Calculate the confidence $c$ using temporal increment along the twig and choose the initial orientation to be the one maximising the confidence.
		\EndFor
		\For{each twig in dynamical\_graph}
			\State Calculate the local score $s$ of the twig using $c$ the confidence and the spatial alignment with its neighboring twigs.
		\EndFor
		\State Maximise the total score $S = \sum s$ using MonteCarlo algorithm to flip the orientation of twigs while prohibiting the creation of sources.
	\end{algorithmic}
\end{algorithm}

\begin{algorithm}[!h]
	\caption{Function: BRANCH\_IDENTIFICATION\\
	Identify each branch according to the orientation of twigs in an oriented\_twigs\_graph}\label{alg:Branchification}
	\begin{algorithmic}[]
		\Require oriented\_twigs\_graph\\
		\State Mark the three twigs going out of the source as the start of new branches
		\For{each degree 3 nodes in oriented\_twigs\_graph}
			\State Identify the three twigs sharing this node. By construction, except for the source, there must be at least one incoming twig and at least one outgoing twig.
			\State Among these three twigs, associate one incoming twig with one outgoing twig according to time synchronisation and use spatial alignment as a tie-breaker if necessary.
			\If{the third twig is an outgoing twig}
				\State Mark this twig as the start of a new branch
			\EndIf
		\EndFor
		\For {each start of branch}
			\Repeat
				\State Prolongate the branch by adding the associated twig at the current end of the branch.
			\Until{No more twigs are available to continue}
		\EndFor
	\end{algorithmic}
\end{algorithm}

\begin{algorithm}[!h]
	\caption{Function: OVERLAPS\_CORRECTION\\
	Identify and correct every overlaps in a branched\_graph}\label{alg:OverlapsCorrection}
	\begin{algorithmic}[]
		\Require branched\_graph\\
		\For{each branch $B$ in branched\_graph}
			\State List all pairs ($B_{in},B_{out}$) where $B_{in/out}$ correspond to any branch incoming in/outgoing from the branch $B$.
			\For{each pair ($B_{in},B_{out}$) in this list}
				\State Note $t_{in}$ the time when the incoming branch $B_{in}$ meets the branch $B$ and $t_{out}$ the instant the branch $B_{out}$ grows out of the branch $B$. We note $\Delta t_o = t_{out}-t_{in}$.
				\State Note $L_o$ the length along $B$ between the junction of $B_{in}$ and $B$ and the junction of $B_{out}$ and $B$.
				\If{$L_o - VT\leq V \Delta t_o \leq L_o + VT$ \textbf{and} $L_o \leq \mathcal{L}$}\Comment{$\mathcal{L}$ is a threshold set to 5 hyphae diameters in our case}
					\State Mark the triplet ($B,B_{in},B_{out}$) as a potential overlap.
				\EndIf
			\EndFor
		\EndFor
		\If{a branch appear more than once as an incoming branch or an outgoing branch}
			\State Keep only the potential overlap with the best spatial alignement between the incoming and the outgoing branches.
		\EndIf\\
		\For{each potential overlap ($B,B_{in},B_{out}$) remaining}
			\State Remove the connection from branch $B_{in}$ to the branch $B$ and from the branch $B$ to the branch $B_{out}$.
			\State Add a new node $n$ at the mean position of the overlap and fuse the overlapping branches into a single branch of the form $B_{in} - n - B_{out}$.
		\EndFor
	\end{algorithmic}
\end{algorithm}

\clearpage
\bibliography{biblio.bib}
\end{document}